\documentclass[aps,pre,twocolumn,showpacs,superscriptaddress]{revtex4}
\usepackage{graphics}
\usepackage{epsfig}
\usepackage{amsmath}
\begin{document}

\title{Navigating networks with limited information}

\author{M. Rosvall}
\email{rosvall@tp.umu.se}
\affiliation{Department of Theoretical Physics, 
Ume{\aa} University, 901 87 Ume{\aa}, Sweden}
\affiliation{NORDITA, Blegdamsvej 17, Dk 2100, Copenhagen, Denmark}
\homepage{www.nordita.dk/research/complex}
\author{P. Minnhagen}
\affiliation{Department of Theoretical Physics, Ume{\aa} University, 901 87 Ume{\aa}, Sweden}
\affiliation{NORDITA, Blegdamsvej 17, Dk 2100, Copenhagen, Denmark}
\homepage{www.nordita.dk/research/complex}
\author{K. Sneppen}
\affiliation{NORDITA, Blegdamsvej 17, Dk 2100, Copenhagen, Denmark}
\homepage{www.nordita.dk/research/complex}

\date{\today}

\begin{abstract}
We study navigation with limited information in networks
and demonstrate that many real-world networks have a structure which can be described as favoring 
communication at short distance at the cost of constraining 
communication at long distance. 
This feature, which is robust and more evident with limited than with complete information,
reflects both topological and possibly functional design characteristics.
For example, the characteristics of the networks studied derived from a city
and from the Internet are manifested through modular network designs.
We also observe that directed navigation in typical networks
requires remarkably little information on the level of individual nodes.
By studying navigation, or specific signaling, we take a complementary approach
to the common studies of information transfer devoted to broadcasting of information
in studies of virus spreading and the like.
\end{abstract}
\maketitle

\section{Introduction}
The study of networks is one possible way to address
the relative importance or ease of communication ability in complex systems
\cite{friedkin,rosvall}. 
In this context a large effort has been devoted to the non specific
broadcasting that dominates for example 
the Internet in the form of spam mail or computer viruses
\cite{moreno,newmanVirus}.  
Here we instead focus on specific signaling
since it has been suggested that that sending a signal to
one specific node without disturbing the remaining 
network is a possible candidate for a design principle
in real-world networks \cite{city}.
By introducing the search information, we have addressed 
this in a general framework in \cite{hide-seek,horizon}
and in relation to urban organization in \cite{city}.
The philosophy of specific communication between a source node
$s$ and a target node $t$ is that $s$ can only send one signal 
that subsequently has to be directed to the 
desired target node $t$.
In principle, for a connected complex network any target $t$
can be reached from any other node $s$, but distant communication is obviously
neither as easy nor as accurate as close direct communication \cite{friedkin}.
In particular, for the social networks studied by \cite{friedkin}
it was observed that knowledge of other people's 
activities declined exponentially with their 
separation in a network and increased linearly with
the number of degenerate paths between them. 

To capture this observation, we here use walks in networks.
The simplest walker is the random walker, 
which has earlier been used to 
characterize topological features of networks \cite{bilke,monasson}, 
including first-passage times \cite{noh},
large-scale modular features \cite{eriksen}, and
search utilizing topological features \cite{adamic}.
Using a simple extension of a random walker, we here discuss
navigation in complex networks.

We consider a random walker that represents the propagating signal released from $s$.
Its probability to reach node $t$ before getting lost on some 
nondirect and thus nonspecific path 
is ${\cal P} \propto \sum_{\{\mathrm{path}\}} \prod_{j \in \mathrm{path}(s,t)} 1/k_j$.
Here the sum captures the linear gain in alternative shortest paths
and the product represents the exponential decline in 
probability as distance increases, thus reflecting the overall
functional dependence observed in \cite{friedkin}.
In real networks the decline of signals over a node
may be faster than $1/k$, representing the possibility that signals are lost.
In our approach we neglect such losses and turn the 
probability to reach a specific node to the minimal 
information 
$I=\log_2({\cal P})$ 
necessary 
to travel directly between the two nodes.
Thereby, \cite{hide-seek,city} characterized networks in terms of the minimal information 
needed to send walkers between specific nodes.

In this paper we implement the specific signaling by letting a walker move from 
source $s$ to target $t$ and make choices of exit links along the walk.
This choice is at every node associated to a node information cost that goes up with the degree of the node.
For a single correct exit among $k$ edges, the minimal information cost would be $\log_2(k)$ bits.
One could easily imagine a higher cost, but we here limit our investigations to the  
optimal organization of information at each node.
The total information cost in going from a source to a target is counted as the sum 
of all the individual node information costs along the way. 
As high-degree nodes cost more to pass than low degree ones, we 
find that the total information cost depends crucially on the 
relative organization of high- and low-degree nodes, 
as well as on modular features of the network.

In practice, the walk from $s$ to $t$ may be more
or less directed, dependent on the walkers ability to choose exit links that
lead it closer to the target.
If the walker at each point along the walk chooses an exit link $e$
that leads it closer to the target, it will arrive to the target node $t$
after $l_{st}$ steps equal to the shortest path between $s$ and $t$.
But if the access to node information is ``limited" along the way, there are chances to make mistakes:
The walker has a probability to choose an edge that increases its distance to the target.
The length of the walk will then be longer---say, with an amount $\Delta l_{st}$ compared with the shortest path $l_{st}$.
The total information cost is then determined by two factors: 
directly by the limited node information $\imath$ and indirectly by the length of the path, which 
increases with decreased node information.
As the limits on node information $\imath\rightarrow 0$ the walk will be random.
The limits on the node information affect nodes of high degree 
more than low-degree ones; hence the structure of the underlying network plays an
important role in the interplay between typical walk length and the limited node information. 

The information measure presented is interesting for two reasons:
First it captures the information cost payed by a ``signal'' in some
real world scenarios---for example, a newcomer in a city asking about the way to
the hotel \cite{city}---where the limited information approach corresponds
to not asking enough questions and instead to high or low extent walk by chance.
Second it provides a method to characterize and distinguish networks 
qualitatively and quantitatively from each other.

%the total information used to make
%he walk will be a non-monotonous function of the access to node information.

\section{Search Information}

We first quantify the information cost in number of bits $I(s \to t)$
it takes to navigate the shortest path from
node $s$ to node $t$. This could in principle be done as in Ref.\ \cite{hide-seek}
but given that we have to compute $I(s \to t)$ on the basis of local choices
we compute a node information $\imath_{jt}$ on every node $j$
on a walk leading to target $t$. That is, $\imath_{jt}$ is the number of
bits that one needs on node $j$ in order to select one of the exits that leads to $t$
along a shortest path.
Then, following the walker from $s$ to $t$ we compute
$I(s \to t) = \sum_{j \in \mathrm{path}(st)} \imath_{jt}$.

\begin{figure}[hbt]
\centering
\includegraphics[width=\columnwidth]{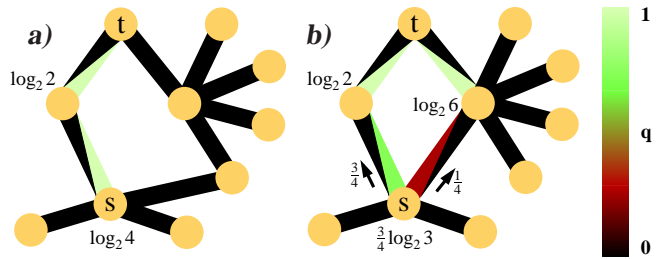}
\caption[] {(Color) Search information \mbox{$I(s \to t)$} measures
the average number of bits one needs in order to walk
along one of the shortest path from $s$ to $t$. It can be sampled by
walking along the shortest paths between $s$ and $t$: (a) For nodes with a unique direction to the target,
this direction is selected with an information cost $\log_2(k_j)$.
(b) For nodes $j$ at branch points between degenerate paths, the exit links
are selected with probability $q_{jit}$ (colored after value) proportional to
$p_{jit}$, the probability that a random walker at that exit would
reach the target along a shortest path. 
Let, for example, the probability to leave
node $s$ to the left be $q$ and accordingly
$1-q$ to take the right way (the probability is zero to go downwards
in the figure since these links are not on a shortest path to $t$).
Then, $I(s \to t)(q) = [\log_2 4 + q\log_2 q + (1-q)\log_2 (1-q)]
+ [q\log_2 2 + (1-q)\log_2 6]$,
where the first set of parentheses is the information cost on node $s$ and second set of parentheses
the cost on the following step, left and right, respectively.
We look for a $q$ that minimizes $\{I(s \to t)(q)\}$, giving
$q=(1/2)/(1/6+1/2) =0.75$. \mbox{$I(s \to t) = 3.0$} in
(a) and \mbox{$I(s \to t) \approx 2.6$} in (b).
} \label{fig1}
\end{figure}

If no degenerate paths exist, as in Fig.\ \ref{fig1}(a), then
\begin{equation}
\imath_{jt} = \log_2 k_j,\label{nodeg}
\end{equation}
where $k_j$ is the degree (number of links)
of node $j$, since the task is to select one link among $k_j$.
$\imath_{jt}$ can also be understood as the information loss
associated with weighting all links equally, instead of
knowing the unique exit path.
When there are two or more degenerate paths from $j$ to $t$,
the required information depends on the relative probabilities
that one wants to choose each shortest paths with, and
Eq.\ (\ref{nodeg}) generalizes to
\begin{equation}
\imath_{jt}=\log_2(k_j)+\sum_{i}q_{jit}\log_2 q_{jit},
\label{degen}
\end{equation}
where $q_{jit}$ is the
probability to choose a link to node $i$ from node
$j$ on a walk to node $t$ ($\sum_{i}q_{jit}=1$).
$q_{jit}=0$ if the link is not on the shortest path between $j$ and $t$
(or if there is no link between $j$ and $i$).
Equation (\ref{degen}) counts the information loss associated with setting all $k_j$ links
equal instead of confining them with the selection probabilities
$q_{jit}$. Thus it also counts the information needed
to confine our choice to the limits imposed by $q_{jit}$, given that
one has to choose one of $k_j$ exit links.
For example, if all paths are degenerate and chosen with equal weights, $\imath_{jt}=0$,
whereas two equally weighted degenerate paths would contribute with
$\imath_{jt}=\log_2(k_j)-\log_2(2)$.
Following the line to always choose the method or parametrization that represents
the lower limit of information we choose the probability to leave a node along a link on a shortest path
between $s$ and $t$ to minimize the total information
cost $I(s \to t)$:

In general, if there are many degenerate paths, the probability to exit
to node $e$ from node $j$ on the shortest path to $t$ is
\begin{align}\label{qvalues}
q_{jet}=\frac{p_{jet}}{\sum_{i} p_{jit}},
\end{align}
where
\begin{align}\nonumber
p_{jit}= \sum_{\mathrm{path}(it)}
\prod_{l \in \mathrm{path}(jit)} \frac{1}{k_l}.
\end{align}
$ p_{jit}$ is the
probability to walk the shortest path to $t$
from node $j$ via the link to node $i$ in an unbiased walk.
For example, in  Fig.\ \ref{fig1}(b), $p_{sit}$ is $1/2$ to the left
and $1/6$ to the right.
Thus, not all degenerate paths are weighted equally: It pays off to use
additional information on a node with branching paths, in order to avoid paying more
information later. In this way one suppresses paths which goes through nodes that
have high degrees and thus are
more costly to pass [e.g.\ right path in Fig.\ \ref{fig1}(b)].
We note that the results from the particular choice of $q$ values
are chosen to minimize the total information cost as we want to be consistent
with our overall optimal search approach.
In practice this biased branching gives bits of search information which are
almost inseparable from the search where each correct link is chosen with
equal probability.

For each walk, $I(s\to t)$ is the sum
of the contributions along the shortest path from source $s$
to $t$. If there are degenerate paths, $I(s\to t)$ is calculated by averaging over many walks
from $s$ to $t$. In this way we by local walks obtain the search information
for any pairs of nodes $s$, $t$, an information that may also be calculated
directly \cite{hide-seek} from knowing all degenerate paths $\mathrm{path}(st)$ between $s$ and $t$:
\begin{equation}
I_S(s\rightarrow t) = - \log_2\left( \sum_{p(s,t)}  \prod_{j\in \mathrm{path}(st)} \frac{1}{k_j} \right).
\end{equation}
The present definition differs slightly from that of \cite{hide-seek} because we here
are also open to the possibility of returning to a node that just was visited 
[$1/k_j$ instead of $1/k_j-1$].
Here we do not exclude a step back, since
we want to generalize our measure to the case of nonperfect walks associated with limited
or imperfect node information.

\section{Limited Search Information}

\begin{figure}[hbt]
\centering
\includegraphics[width=\columnwidth]{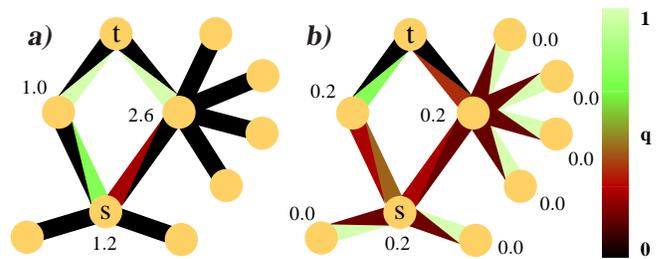}
\caption[] {(Color) Search with limited information: At each node
$j$ along a path from $s$ towards the target $t$,
the information cost is limited by $\imath$ bits, or equivalently,
only $\imath$ bits are accessible.
The color of the links represents $q_{jit}$, the probability to leave node $j$
along a link to node $i$ on a path towards $t$.
With limited information the weighted exit probabilities $q_{jit}$ are changed according to
$(q_{jit}+\epsilon_{jt})/(1+k_j \epsilon_{jt}) \to q_{jit}$
to satisfy $\imath_{jt} \le \imath$. In (a)
there is no upper limit, or $\imath=\infty$, and the case reduces to
the one in Fig.\ \ref{fig1}(b) with $I_\infty(s \to t) \approx 2.6$ bits
and the excess walk $\Delta l = 0$.
In (b) $\imath=0.2$ and $I_{0.2}(s \to t) \approx 0.7$ bits
and the excess walk $\Delta l \approx 2.5$.
} \label{fig2}
\end{figure}

We now turn to the limited information perspective
and assume that the
amount of information at a node is limited to $\imath$ bits (illustrated in Java applet \cite{java}).
The consequence is that the walker increases its probability
to not follow a shortest path as $\imath$ decreases. Further,
the walk between, say, $s$ and $t$ can be substantially longer than
the actual shortest path. In Fig.\ \ref{fig2}(b), $\imath = 0.2$ and the average
walk is 4.5 steps compared to the 2 steps in Fig.\ \ref{fig2}(a).
To limit $\imath_{jt}$ to $\imath$ we blur the $q$ values of node $j$ in Eq.\ (\ref{qvalues}) 
by a $\epsilon_{jt} \in [0,\infty]$,
through a uniform smearing $q_{jit} \rightarrow q_{jit}+\epsilon_{jt}$
that increases the probability to choose any false exit with an equal amount.
Normalizing the local exit probabilities we obtain the smeared
exit probability
\begin{equation}
q_{jit} \rightarrow q_{jit}(\epsilon_{jt}) \; =\;
\frac{q_{jit}+\epsilon_{jt}}{1+k_j \epsilon_{jt}} ,\;
\end{equation}
which interpolates between the optimal
nonblurred value $q_{jit}$ and the
random walk value $1/k_j$ in a simple way.
$\epsilon_{jt}$ is determined to satisfy
\begin{align}
\imath_{jt}=\log_2(k_j)+\sum_{i}q_{jit}\log_2 q_{jit} \le \imath ,
\end{align}
where $<$ is only relevant when the unblurred $\imath_{jt}$ is already lower than 
the limited node information $\imath$.
The effect of limited information
on choosing a unique correct exit link varies with the degree $k$ of a node.
With information threshold $\imath=1$
the probability assigned to this single correct exit
link is $17\%$ if $k=1000$, $28\%$ if $k=100$,
$68\%$ if $k=10$, and obviously $100\%$ if $k \le 2$.\\

In order to quantify the information associated with walking
in different environments (networks) and information limits $\imath$ we 
consider the average number of bits of information it takes to
navigate in the network with $N$ nodes,
$I_{\imath} = 1/N^2\sum_{s,t}I(s \to t)$.
$I_{\imath}$ thus quantifies the navigability
or search information of networks as in Ref.\ \cite{hide-seek},
but takes into account the limited node information and associated usage of
nonshortest paths.

\section{Results and Discussion}

\begin{figure}[hbt]
\centering
\includegraphics[width=\columnwidth]{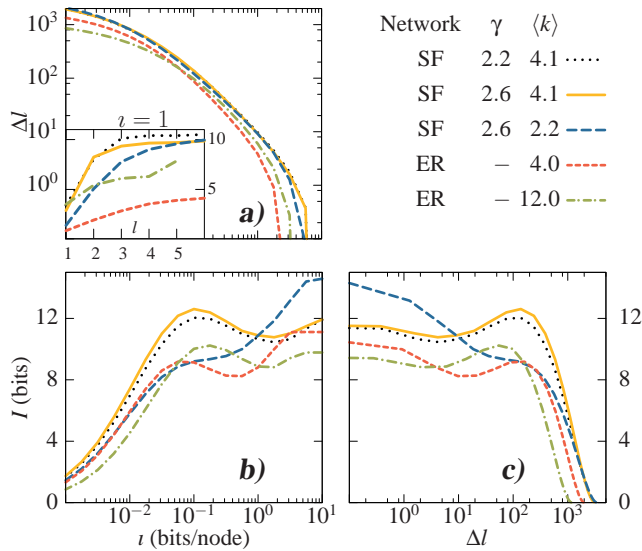}
\caption[] {(Color online) Search with limited information in
Erd{\H o}s-R\'enyi (ER) networks and scale-free (SF) networks with degree
distribution $P(k)\propto 1/(k_0+k)^{\gamma}$ parametrized by
$\gamma$ and $\langle k \rangle$. (a) As the available
information at each node decreases with decreasing $\imath$, the typical path
length for going between two nodes increases by
$\Delta l$ beyond the length of the shortest path between the nodes. The
inset demonstrates that $\Delta l$ is nearly independent of the
length of the shortest path between nodes for $l>2$.
(b) The variation of $I_{\imath}$ as a function
of the available node information $\imath$.
(c) The typical
search information $I_{\imath}$ first decreases and then subsequently
increases as the walk length increases due to the
limits on $\imath$.}
\label{fig3}
\end{figure}

Figure \ref{fig3} shows the effects a limited $\imath$ has on
a number of model networks.
All networks have $10^4$ nodes and are two Erd{\H o}s-R\'enyi (ER)
networks with two different average degrees and two scale-free (SF) networks with degree
distribution $P(k)\propto 1/(k_0+k)^{\gamma}$ parametrized by
$\gamma$ and $\langle k \rangle$.
With this parametrization it is possible 
to keep the same number of links in the two scale-free networks with
different exponents. The networks are generated by the method presented in \cite{hierarchy},
which ensures that they are uncorrelated and connected.
Overall $I_{\imath}$ is nonmonotonous in $\imath$.
For very high $\imath$ ($>10$ in Fig.\ \ref{fig3}) one reproduces
the search information of a direct walk.
As $\imath$ is decreased to, say, $\imath \sim 1$,
the total search information is decreased by a few bits,
reflecting the fact that the
gain by asking fewer questions at each node is slightly larger
than the cost of going a few steps longer. In fact, the local minimum roughly corresponds
to walks which typically are $\Delta l\sim 10$ steps
longer than the direct path [Figs.\ \ref{fig3}(a) and \ref{fig3}(b), 
a length comparable to the diameter of the networks.
This $\Delta l$ is representative for typical paths, independent
of the direct distance $l$ (for $l>2$) between source and target, as illustrated
in the inset of Fig.\ \ref{fig3}(a).

For even smaller $\imath$ the rapidly
increasing length of the walks makes the the total information cost
$I_{\imath}$ increase; for some networks
it even becomes larger than $I_{\imath=\infty}$.
For still smaller $\imath$ the walk gradually approaches that of a random
walk and the length of the walk is limited by system size.
Thus for small enough $\imath$ the $I_{\imath}$ is bound to decrease to zero,
a decline that starts for $\imath$ decreasing to values below $\sim 0.1$ bit.

The maximum of $I_{\imath}$ obtained around $\imath\sim 0.1$ bit is most striking for
networks with high average degree and especially if they have broad
degree distributions. That is because the information constraint is
strongest on the high-degree nodes, where one in
principle needs more information to navigate correctly.
To investigate this further we have examined
walks of the type $s\rightarrow s$---i.e.\ walks that start and end in the same node.
Roughly independent of the investigated network, we found that as
$\imath$ decreases below 1 bit, the walks starts to delocalize
and are completely delocalized at $\imath\sim 0.1$. This corresponds
to the information threshold at which
the walk lengths depicted in Fig.\ \ref{fig3}(a) start to saturate
and $I_{\imath}$ reaches a maximum.
Obviously the value of $\imath$ where the walkers localize
increases with the average degree $\langle k\rangle$.

\begin{figure}[hbt]
\centering
\includegraphics[width=\columnwidth]{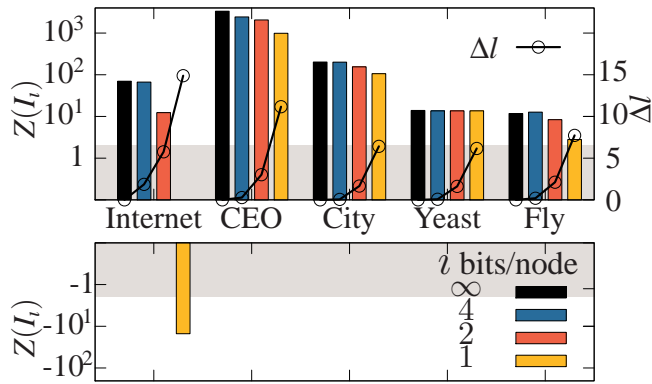}
\caption[] {(Color online) Overall navigability of real-world networks,
compared to their random counterparts presented as $Z$ scores, for four
levels of $\imath=\infty,4,2,1$ bits/node together with the
corresponding average excess path length $\Delta l$. $\Delta l$ shown in the figure
is for the real networks, but it also very well represents the randomized counterparts. 
The Internet (hardwired Internet
of autonomous systems \cite{internet}) is more sensitive to limited information
than the similarly sized CEO (chief executive officers connected by links if
they sit at the same board \cite{ceo}).
The city network is the Swedish city Malm{\"o} with streets mapped to nodes and
intersections mapped to links \cite{city}. The two biological betworks
are the protein-protein interaction networks of \emph{Saccharomyces Cerevisiae}
(yeast) \cite{Uetz2000}
and \emph{Drosophilia melanogaster} (fly) \cite{giot}.
} \label{fig5}
\end{figure}

The navigability of a network is determined by its topology.
That is, it depends on both the degree distribution and how
nodes of various degrees are connected to each other.
We will here focus on comparing a given real-world network with
its randomized counterparts, defined
by rewiring links such that all nodes conserve their degree and such that the
network remains globally connected \cite{maslov2002}.
To quantify navigability in the presence of limited information
we compare the Z scores
[$Z(I_{\imath})=(I_{\imath}- I_{\imath}^{\mathrm{random}} )
/\sigma_{\imath}^{\mathrm{random}}$], where $I_{\imath}^{\mathrm{random}}$
is the average $I_{\imath}$ for corresponding randomized networks,
with standard deviation
$\sigma_{\imath}^{\mathrm{random}}$.
In Fig.\ \ref{fig5} we
investigate real networks at four different levels of limited information.

The Internet is the network of autonomous systems \cite{internet} that 
in this data-set consists of 6474 nodes and 12 572 links and its degree 
distribution is scale free with $P(k)\propto 1/k^{2.1}$.
In the CEO network (6193 nodes and 43 074 links),
chief executive officers are connected by links if
they sit at the same board \cite{ceo}.
The city network is constructed by mapping 1868 streets to nodes and
3026 intersections to links between the nodes in the Swedish city Malm{\"o} \cite{city}.
Yeast is the protein interaction network
in \emph{Saccharomyces Cerevisia} 
detected by the two-hybrid experiment \cite{Uetz2000},
and fly refers to the similar network
in \emph{Drosophilia melanogaster} \cite{giot}.
Both of these networks are pruned to include only interactions
of high confidence, and in both networks we compare with their
random counterparts where both bait and prey connectivity of all proteins
are preserved.

Overall, all networks but the Internet maintain their rather bad navigability
with decreasing node information. Thus the overall communication features
reported earlier \cite{hide-seek} are robust to the limited information
and searches that go beyond the shortest path.
The particular result of the Internet means that its randomized versions
are more difficult to navigate with low information.
%Internet is more sensitive since it 1) is known to be modular \cite{eriksen},
%and 2) has a broader degree distribution
%with nodes up to degree $\sim 1000$.

\begin{figure}[hbt]
\centering
\includegraphics[width=\columnwidth]{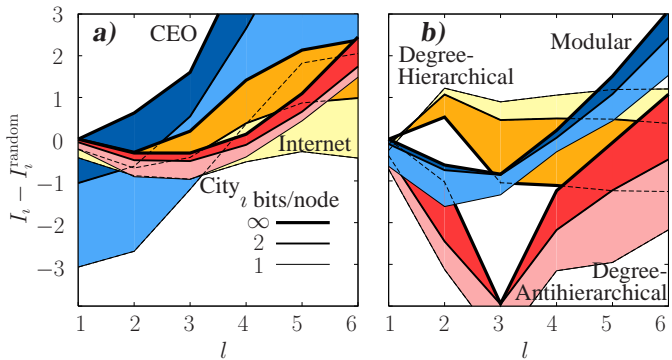}
\caption[] {(Color) Information horizon of (a) 3 real-world
networks from Fig.\ \ref{fig5} and (b) three model networks
of size $N=1000$; one modular and two
scale-free networks with degree distribution $P(k) \propto k^{-2.4}$,
organized to be, respectively, degree hierarchy and degree antihierarchy
\protect{\cite{hierarchy}}.
We compare information associated to navigation between nodes at distance $l$, with
the navigation in randomized counterparts (keeping the degree sequence) for
$\imath=\infty, \; 2$, and $1$.
}
\label{fig6}
\end{figure}

To understand the overall navigability 
in more detail, Fig.\ \ref{fig6} resolves $I_{\imath}$
into $I_{\imath}(l)$, defined as the average search information over
all nodes separated by a shortest path length $l$.
We examine the average information associated to 
walking to a specific node a distance
$l$ away in the network \cite{horizon} and include three
model networks with $1000$ nodes that show distinguishing features.
The modular network is constructed by 33 highly
interconnected communities, each node having 6 links to nodes within the community
and each community having 6 links to other communities.
The degree hierarchical network is constructed so that
all shortest paths have the property that they first go
to nodes with subsequently higher degree (up in the degree hierarchy) and then
to nodes with lower and lower degree to the target.
In the degree antihierarchy the networks are constructed to
minimize this property \cite{hierarchy}.
In order to renormalize for effects associated with the degree distribution
we also here compare with the corresponding $I_{\imath}^{random}(l)$
for the randomized counterparts.

Let us as an example discuss the Internet, where
$I_{\imath}-I_{\imath}^{random}(l)$ exhibits
a minimum for $l\sim 2\rightarrow 3$ at all $\imath \ge 1$.
This reflects a modular structure associated to country boundaries \cite{eriksen}.
Walks within the modules visit highly connected nodes
less frequently than in the randomized version where even short
paths tend to go through the hubs. 
In contrast, when forcing paths to go 
through highly connected nodes at very short distances, as they do in 
the degree hierarchy [see Fig.\ \ref{fig6}(b)], $I_{\imath}(l\sim 2)$
becomes relatively large at short distances.

\begin{figure}[hbt]
\centering
\includegraphics[width=\columnwidth]{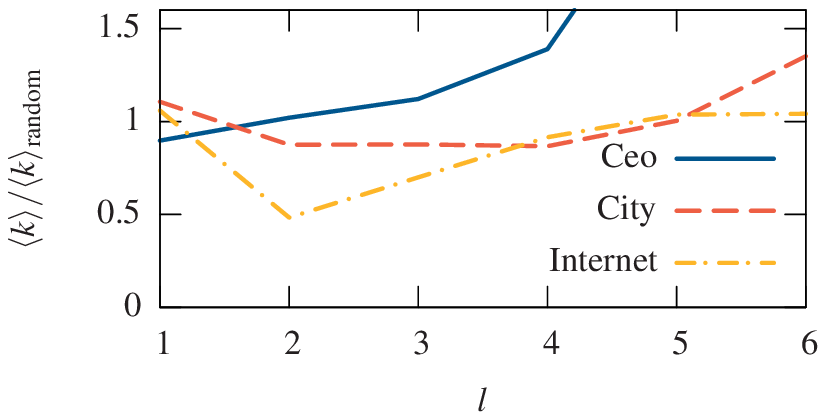}
\caption[] {(Color online) The average degree $\langle k \rangle$ of nodes as function
of distance from a random node, in units of 
what it is in a randomized version. The networks are the same as studied in Fig.\ \protect{\ref{fig6}}(a).
A relative high value of $\langle k \rangle$ is associated with the information barrier,
as indeed seen by comparing $\langle k({\ell}) \rangle / \langle k({\ell}) \rangle_{random}$ 
with the $I({\ell})-I_{random}({\ell})$ in Fig.\ \protect{\ref{fig6}}(a).
}
\label{fig7}
\end{figure}

For distances $l>3$ the path typically
escapes the module and goes through highly connected nodes.
The advantage of modular structures at short distance is turned to a disadvantage
at long distance, as is also illustrated for the modular network in Fig.\ \ref{fig6}(b).
As a consequence, also for the Internet,
$I_{\imath=\infty}-I_{\imath=\infty}^{random}(l)$ is positive at large $l$.
A possible interpretation is that at these larger distances, the country modules are connected through
nodes of high degree, as reflected by the overabundance of high-degree nodes
at distances ${\ell}\sim 4$ in Fig.\ \ref{fig7}.
Thus modular structure connected by high-degree nodes
gives the Internet an information horizon at these intermediate distances.

With information limited to $\imath$, the information cost at especially the
highly connected nodes reduces extensively.
This is especially important for the Internet, where the chance to make 
mistakes on the many large hubs increases substantially with decreased $\imath$.
In fact, with decreasing $\imath$ the search gets cheaper in the Internet, but
more expensive in its randomized counterparts. 
This is because walks in the randomized version bounce between the hubs, 
which are more interconnected than in the real Internet \cite{maslovInternet}. 
This bouncing adds to the total information cost by the high cost to
pass by hubs.
In contrast, as $\imath$ decreases the real Internet in fact increases its communication
ability because many of the false exits lead to nodes of degree 1
where the walker bounces back without information cost.
Figure \ref{fig6}(b) reveals a similar communication topology in the
degree antihierarchy.

For the CEO network, the most striking pattern is that limited information walks
at short distance are much easier in the real, than in its randomized counterparts.
The walks are quite localized in the CEO network,
a direct consequence of the highly modular structure of the
fully connected boards. The pattern for the city uncovers a modular structure
indicated by the high resemblance with the modular network in Fig.\ \ref{fig6}(b).
The design of this city makes navigation at short distance easier than in a random city
and this feature is even more evident in the perspective of the limit information
indicated by the stronger horizon as the node information is decreased.

\section{Summary}

The design of network topologies defines the ability to direct signals,
thus maintaining cooperativity in the corresponding system.
In this paper we have investigated how the peer-to-peer communication
of networks can be maintained in view of sending signals with the possibility
to make erroneous choices along the signaling paths.
This was done by introducing a walker on the network and quantifies how well
this walker located a given target node, provided more or less correct information
on directions as the walker moved from node to node in the network.
Overall we have found that the results for unlimited node information presented
both in \cite{hide-seek} and \cite{horizon} are robust to limited node information and nonshortest paths.
Thus, the approach to characterize networks with shortest paths is a good proxy for
characterizing also communication where mistakes are allowed.
In particular we have demonstrated that real-world networks
as diverse as the Internet, a city network and in fact also molecular networks (data not shown)  
have a structure which can be described as favoring 
communication on short distance at the cost of constraining 
communication on long distance.
There are two aspects of such communication structure, 
a tendency to modular organization, and a tendency to 
constrain signals to certain channels.
The modular network design is a characteristics of both the studied city and the Internet topologies.
The feature associated to certain communication channels 
was investigated in Fig.\ \ref{fig7} where we found 
a structure with paths that consist of sequences of several lowly connected nodes.
The hubs typically interfere with the walker some length down the paths, and at least for the 
Internet the hubs are associated with communication between the modules.

Finally, and more generally, the fact that one manages fairly well with small node information
in all investigated cases, implies that directed navigation in typical networks
requires remarkably little information on the level of individual nodes.

\subsection*{Acknowledgement}
We acknowledge the support of the Swedish Research Council through
Grant Nos.\ 621 2003 6290 and 629 2002 6258.

\vfill\eject

\end{document}